\documentclass[twocolumn,preprintnumbers,amsmath,amssymb,pre]{revtex4}
\usepackage{txfonts}
\usepackage{amsfonts}
\usepackage{amsmath}
\usepackage{graphicx} 
\usepackage{dcolumn}  
\usepackage{bm}
\usepackage{overpic}
\usepackage{color}
\usepackage{multirow}

\begin{document}

\title{Mechanical impulse propagation in a packing of 3D spheres confined at constant pressure}

\date{\today}

\author{Francisco Santibanez$^{1,*}$, Rene Zu\~{n}iga$^1$, Francisco Melo$^2$}

\affiliation{ $^1$ Instituto de F\'isica, Pontificia Universidad Cat\'olica de Valpara\'iso, Av. Brasil 2950, Valpara\'iso, Chile \\$^2$ Departamento de F\'isica, Universidad de Santiago de Chile, Av. Ecuador 3493, Santiago, Chile.}
\email[Corresponding author: ]{francisco.santibanez@pucv.cl}     
\begin{abstract}

Mechanical impulse propagation in granular media depends strongly on the imposed confinement conditions.  In this work, the propagation of sound in a granular packing contained by flexible walls that enable confinement under hydrostatic pressure conditions is investigated.   This configuration also allows the form of the input impulse to be controlled by means of an instrumented impact pendulum.  The main characteristics of mechanical wave propagation are analyzed, and it is found that the wave speed as function of the wave amplitude of the propagating pulse obeys the predictions of the Hertz contact law.  Upon increasing the confinement pressure, a continuous transition from nonlinear to linear propagation is observed. Our results show that in the low-confinement regime, the attenuation increases with an increasing impulse amplitude for nonlinear pulses, whereas it is a weak function of the confinement pressure for linear waves.
\end{abstract}
\maketitle

\section{\label{sec:intro}Introduction}
Acoustic methods have been used for decades to characterize the mechanical properties of the complex systems that are commonly encountered in geological analyses for mining, construction and soil characterization~\cite{borehole_acoustics_DLJohnson2006, velocity_saturated_soils, Bao2012257,soil_mining}, among others. However, the physical mechanisms that govern sound propagation in such systems remain under investigation and are subject to debate.

Sound propagation though granular materials is a key aspect of a myriad of industrial processes.  
Because of the complexity of the contact networks in these materials, it is expected that nonlinearity, confinement and disorder should play important roles in the propagation of mechanical energy through the media.
However, in most applications, it is necessary to adopt various assumptions, such as a constant sound velocity, homogeneous front propagation and an effective medium approximation; these assumptions have been tested and found to result in discrepancies between experimental findings and numerical simulations of model systems~\cite{Effective_medium_fails_Makse1999}.  

Extensive efforts have been devoted to describing the propagation of mechanical energy in granular media. The features of energy propagation in one-, two- and three-dimensional granular systems have been addressed, revealing the importance of the formation of force chains, which has motivated the study of pulse propagation in 1D-aligned spherical grains.  With regard to this topic, a large volume of important literature has been produced, beginning with the theoretical developments of Nesterenko~\cite{dynamics_heterogeneous_nesterenko2001} and the numerical results of S. Sen~\cite{solitonlike_SSen1998,nonlinear_dynamics_SSen1995} and proceeding through the experimental work of several groups~\cite{wang2015attenuation,nesterenko2013dynamics,chiara_granular_chains, chiara_curved_1d, FS_solitarywave_interaction, Job2008506, Job_Melo_Solitary_waves}.

When considering the propagation of linear waves in three-dimensional granular media composed of perfect spheres confined inside a rigid container, a continuum description leads to a power-law dependence of the speed of sound $c$ on the confinement pressure $P_0$ as follows: $c \propto P_0^{1/6}$.  This result is a direct consequence of the Hertz force law for spheres in contact.  However, this scaling is modified by various effects of practical relevance, such as particle asperities ~\cite{nonlinear_pressure_goddard1990} and moisture~\cite{JIA_wet_absorption}.
When the confining pressure is imposed by gravity, the same continuum limit predicts a mirage effect because, as a result of the increase in pressure with depth, sound travels faster at deeper locations, resulting in the upward deflection of the wavefront~\cite{sound_in_sand}.  More detailed analysis indicates that this particular feature leads to  Rayleigh-Hertz waves \cite{Bonneau_rayleigh-hertz}.

The wavelength of the excitation is another parameter that is important to propagation behavior and determines the form of energy transport. If the wavelength is comparable to the size of the individual grains, then the impulse is scattered from contact to contact, giving rise to a diffusive propagation mechanism~\cite{ultrasound_externalstress_xia1999,ultrasound_disordered_Xia2000,experimental_diffusion_page1995,codalike_xia2004}. In this case, the initial impulse is propagated in two different modes: a coherent ballistic pulse of short duration that propagates along a straight line between the emitter and receiver and a scattered wave that is the result of the superposition of the waves traveling along various paths through the contact network. The coherent pulse, also known as the \textit{P-wave}, arrives first at the receptor and is predominantly longitudinal, whereas the scattered wave, or \textit{S-wave}, is primarily attributable to shear components and is highly sensitive to the internal structural configuration of the packing; the latter has been proposed as a mechanism for identifying changes in the contact network of a material~\cite{coda_interferometry_snieder2006, prlEspindola}.
A propagating impulse can also destroy or create weak contacts in the internal structure of a granular medium. These weak contacts are highly nonlinear and can be probed by small-amplitude perturbations~\cite{weackening_acoustic_xia2011}. The propagation of larger amplitude impulses can sometimes destroy susceptible contacts, thereby triggering the spontaneous emission of secondary pulses, which act as point sources inside the material; in a recent work by X. Jia and F. Giacco~\cite{earthquake_triggering_xia2005,Dynamic_AF_SL}, this mechanism has been proposed as a trigger for earthquakes.\\
Meanwhile, in a weakly confined ensemble of spheres, the continuum description and the Hertz law~\cite{nesterenko2001dynamics} together predict a power-law dependence of the pulse velocity $c$ on the impulse amplitude $P$, namely, $c \propto P^{1/6}$, which has been corroborated in recent experimental investigations~\cite{shockwaves_experimental_wildenberg2013}.
In the present work, we study the propagation of a single mechanical impulse in a dense granular packing under controlled confinement conditions.  The granulate is confined at constant pressure by evacuating a cylindrical elastic container.  The incident and propagated signals are captured by suitable detectors.  This enables the measurement of the propagation speed of a pulse in both the low- and high-confinement limits as well as accurate measurement of the pulse attenuation.  It is found that the propagation speed is nearly independent of the pulse amplitude in the high-pressure confinement limit, whereas it obeys a power law that is consistent with the Hertz contact law at a sufficiently low confinement pressure.   Conversely, a ballistic model combined with viscoelastic dissipation at the sphere contacts captures the main features of the attenuation of high-amplitude pulses propagating through the compact medium.   The same mechanisms of dissipation that are introduced in the propagation equations for a relatively highly confined compact medium also satisfactorily capture the attenuation behavior of low-amplitude linear waves. 

\section{\label{sec:setup} Experimental Setup}
\begin{figure}[h!]
	\begin{center}
		\includegraphics[scale=0.2]{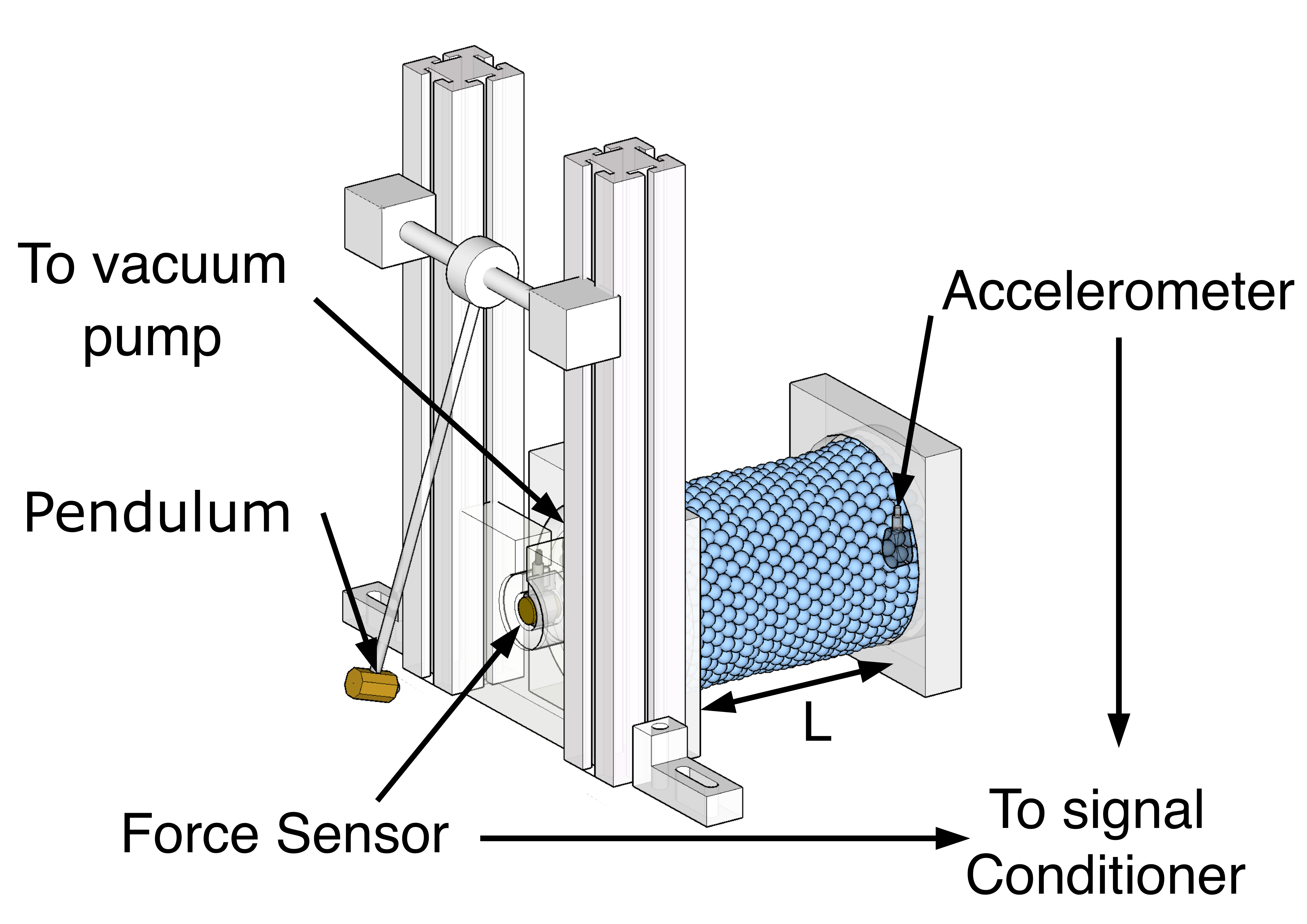}
		\caption{\label{fig:setup} (Color online) Diagram of the experimental setup depicting the impact pendulum and the sensor positioning. For the experiments presented in section~(\ref{subsec:ResonantModes}), the pendulum and force sensor are replaced with a vibration exciter and an accelerometer, respectively.} 
	\end{center}
\end{figure}

In the setup (Fig. \ref{fig:setup}), a packing of approximately $1000$ glass beads (density $\rho_0 = 2400$ kg/m$^3$) of radius $R=2.5$ mm is confined inside an elastic cylinder of length $L=15$cm and diameter $\phi=5$ cm that was formed from a thin latex sheet. The cylinder is hermetically sealed by clamping the latex sheet at each end between two plastic rings that were specially fabricated to allow the emergence of the accelerometer cable and a vacuum hose. A controlled static pressure $P_0$ is imposed on the packing by evacuating the interstitial air using a vacuum pump. This process ensures a constant hydrostatic pressure of as high as $P_0 \approx 93$ kPa and a constant packing fraction of approximately 0.63.   A single short impulse is initiated at one cap of the cylinder by the impact of a rigid pendulum head against a force sensor (PCB Piezotronics model 208C01); the amplitude of the impact is controlled through adjustment of the initial release angle of the pendulum. The impact head of the pendulum consists of a hexagonal brass head with a glass bead mounted on the side that impacts the force sensor; its total mass is $57$ g, and the length of the pendulum is $21$ cm. The impact head was designed to permit modifications to the material and geometry, thereby allowing the duration and form of the input excitation to be controlled. At the opposite end of the cylinder, a miniature accelerometer (PCB Piezotronics model 352A24), located at the center of the cap, records the outgoing pulse. The mass and size of the accelerometer were chosen to be close to those of an individual grain of the medium. \\
The signals from the force sensor and accelerometer are conditioned and amplified by a signal conditioner (PCB Piezotronics model 482C). The outputs of the conditioner and the amplified signal from the pressure sensor (Honeywell 19C015PV5K with an INA114 low-noise amplifier) are acquired by a computer via a simultaneous sampling acquisition card.\\

\section{\label{sec:results} Results} 
\begin{figure}[h!]
	\begin{center}
		\includegraphics[scale=0.25]{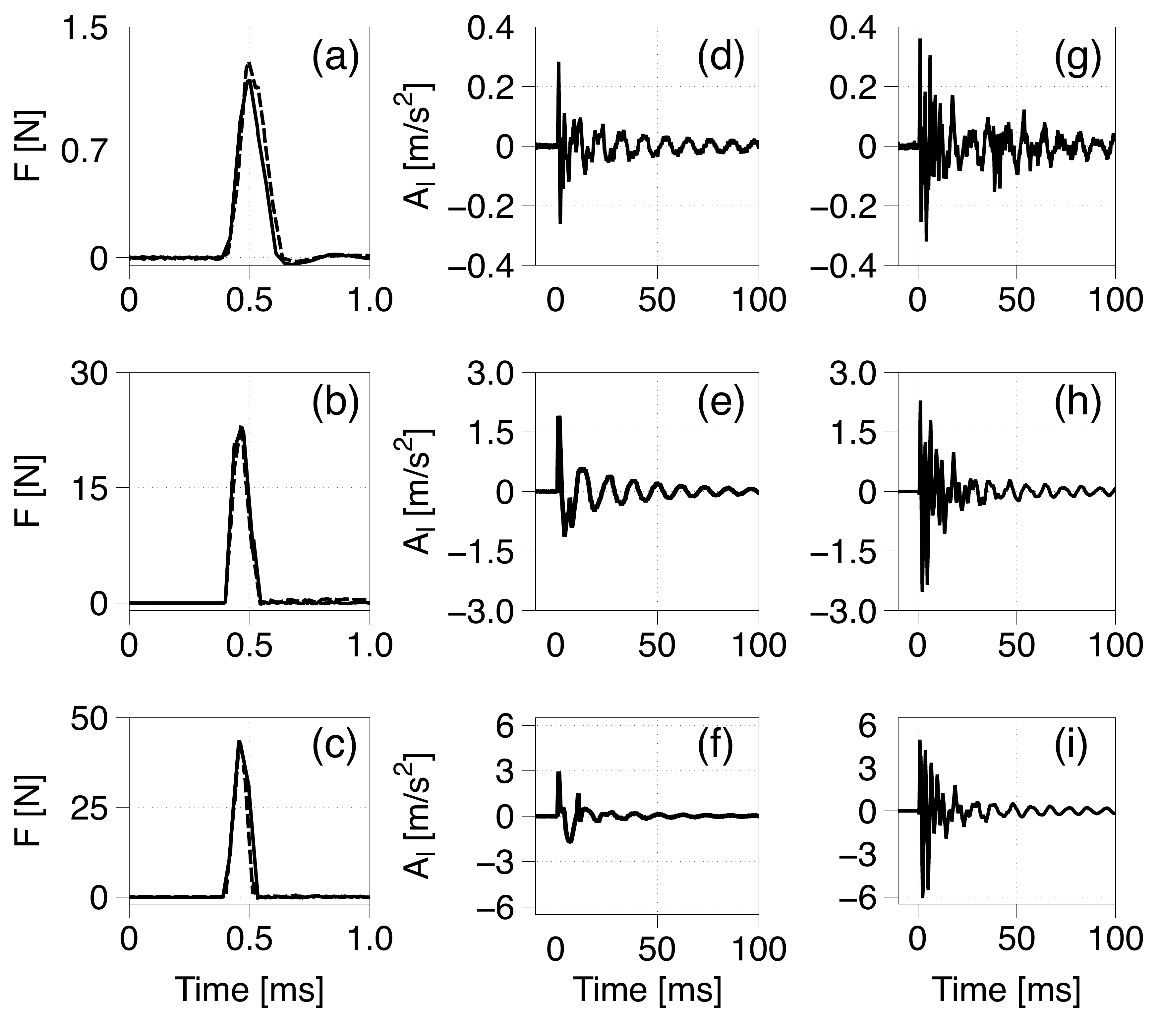}
		\caption{\label{fig:datas} Typical input and propagated signals from the impact and accelerometer sensors, respectively. Left column: input signals.  Middle column: propagated signals for  $P_0=3.2$ kPa. Right column: propagated signals for $P_0 = 83$ kPa.  The rows correspond to different force amplitudes of the incident pulse.  Upper row: $F=1.1$ N. Middle row: $F=22.9$ N. Lower row: $F=41.3$ N.}
\end{center}
\end{figure}

\begin{figure}[h!]
	\begin{center}
		\includegraphics[scale=0.25]{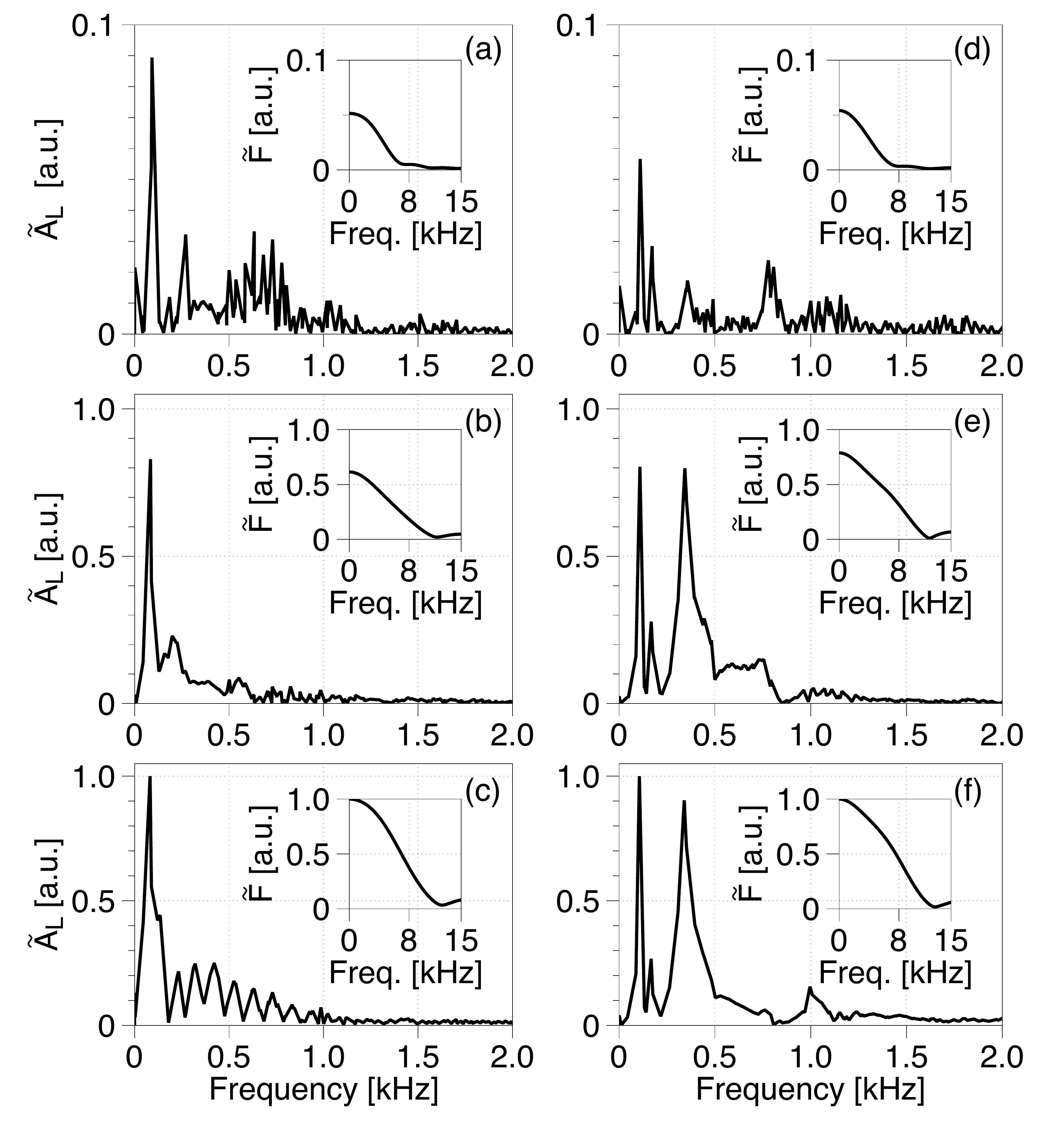}
		\caption{\label{fig:fourier_datas} Fourier transforms of the typical experimental data shown in Fig.~\ref{fig:datas} for the acceleration $\tilde{A}_L$ and force $\tilde{F}$ (in the insets). Each is normalized to the maximum acceleration (or force) amplitude.}
	\end{center}
\end{figure}

Fig.~\ref{fig:datas} shows several typical experimental signals of the input force (left column) and the acceleration at $P_0 = 3.2$ kPa and $P_0=83$ kPa (center and right columns, respectively). The input pulse is much shorter than the output waveform; after the passage of the main pulse, an oscillation remains in the material. This oscillation is more important for the loose packing (low confinement pressure) than for the compact one and is relatively narrow in band. \\
 To investigate the dynamic features of the impulse propagation, we extract the frequency content of the initial and propagated signals by taking their corresponding Fourier transforms (Fig.~\ref{fig:fourier_datas}).  The amplitude of the Fourier transform of the incident pulse (inset of Fig.~\ref{fig:fourier_datas}) indicates that its frequency content extends up to approximately 8 kHz, whereas that of the propagated signal barely surpasses 1 kHz.  \\
At a low confinement pressure, the transmitted wave exhibits a relatively well-defined peak at a frequency of approximately $100$ Hz and a diffuse band up to 1 kHz.  At a high confinement pressure and high excitation, an additional relatively broad peak located near $400$ Hz develops.

\subsection{\label{subsec:Propagation} Pulse Propagation Speed}
We begin by characterizing the pulse speed as a function of relevant parameters.
The maximum of the cross-correlation between the pulse generated as a point source and the first arriving pulse of the acceleration at the sample end (Fig.~\ref{fig:datas}) yields the time delay (flight time) between these signals, which enables the measurement of the propagation speed.  We can then 
study the dependence of this speed on the amplitude of the impact for various confinement pressures. 
 Each measurement is repeated three times, and the average is recorded; then, the same procedure is repeated for a different pressure $P_0$.\\
\begin{figure}[h]
	\begin{center}
		\includegraphics[scale=0.26]{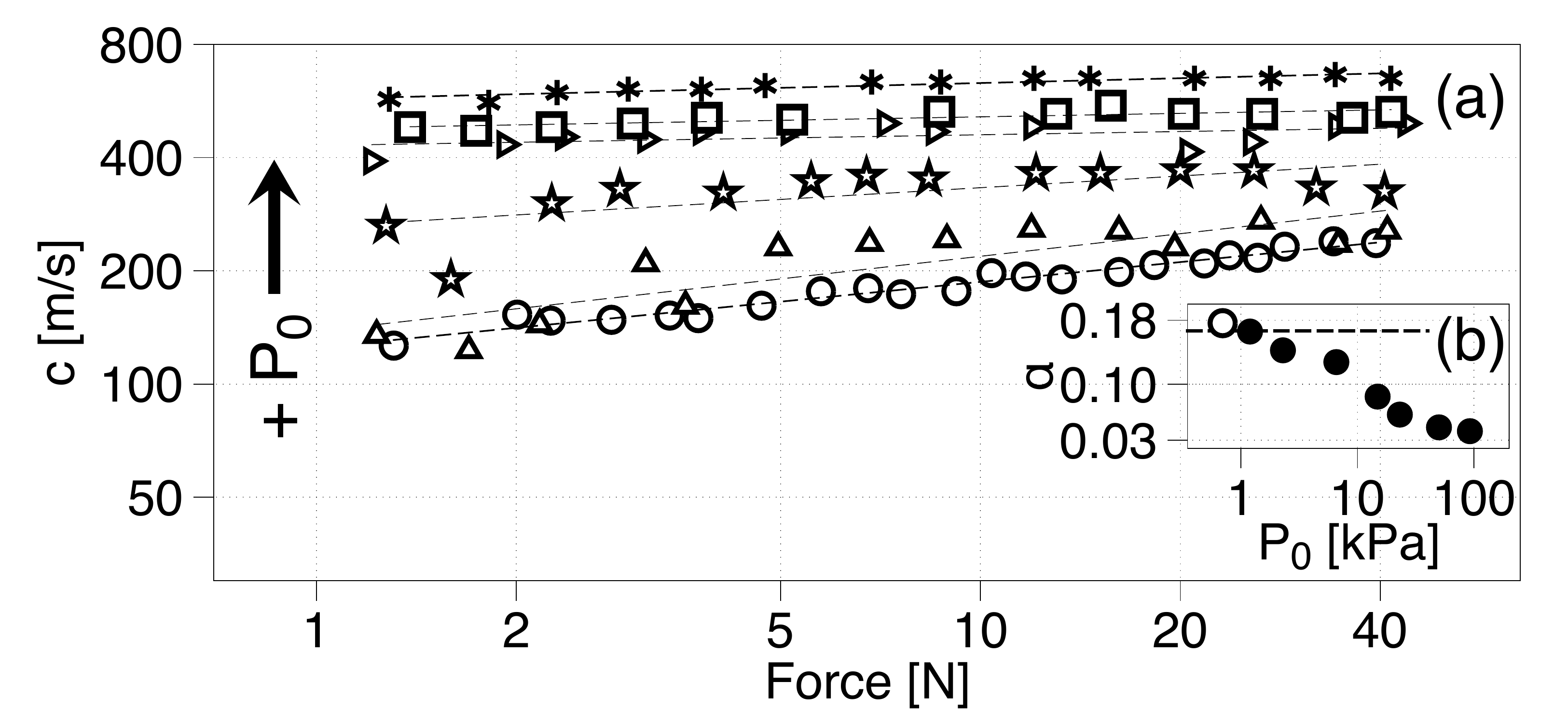}
		\caption{\label{fig:result_wave_speed} (a) Impulse velocity as a function of the excitation force amplitude for $\circ$: $P_0 =0.5$ kPa, $\triangle$: $P_0 =1.2$ kPa, $\star$: $P_0 =14.9$ kPa, $\triangleright$: $P_0 =23.1$ kPa, $\square$: $P_0=50.0$ kPa and $*$: $P_0=92.0$ kPa. Insert (b) shows the exponent $\alpha$ from the fit to $c = C_0 F^{\alpha}$.}
	\end{center}
\end{figure}
In the logarithmic representation, it is apparent that the data follow a power law and that for each data set, the exponent changes for different values of $P_0$, decreasing as the pressure difference increases. The exponents of fits to a power law, $c = C_0 F^{\alpha}$, are shown in Fig.~\ref{fig:result_wave_speed}(b).\\  
We observe a continuous variation from the known $1/6$ exponent~\cite{dynamics_heterogeneous_nesterenko2001,Job_Melo_Solitary_waves,osvanni_2014} to a near-zero value for a high confinement pressure. These findings indicate that the medium undergoes a transition from a nonlinear propagation regime to a linear propagation regime, where the propagation velocity does not depend on the amplitude of the impulse~\cite{shocks_near_jamming}.  The crossover in force associated with this transition has been well characterized for front propagation at a relatively low confinement pressure; see Fig.~3 in  \cite{shockwaves_experimental_wildenberg2013}.  In our experiments, the range of impulse amplitudes is only two decades, which limits the full observation of the crossover. 

\subsection{\label{subsec:ResonantModes} Resonant modes}
\begin{figure}[h!]
	\begin{center}
		\includegraphics[scale=0.28]{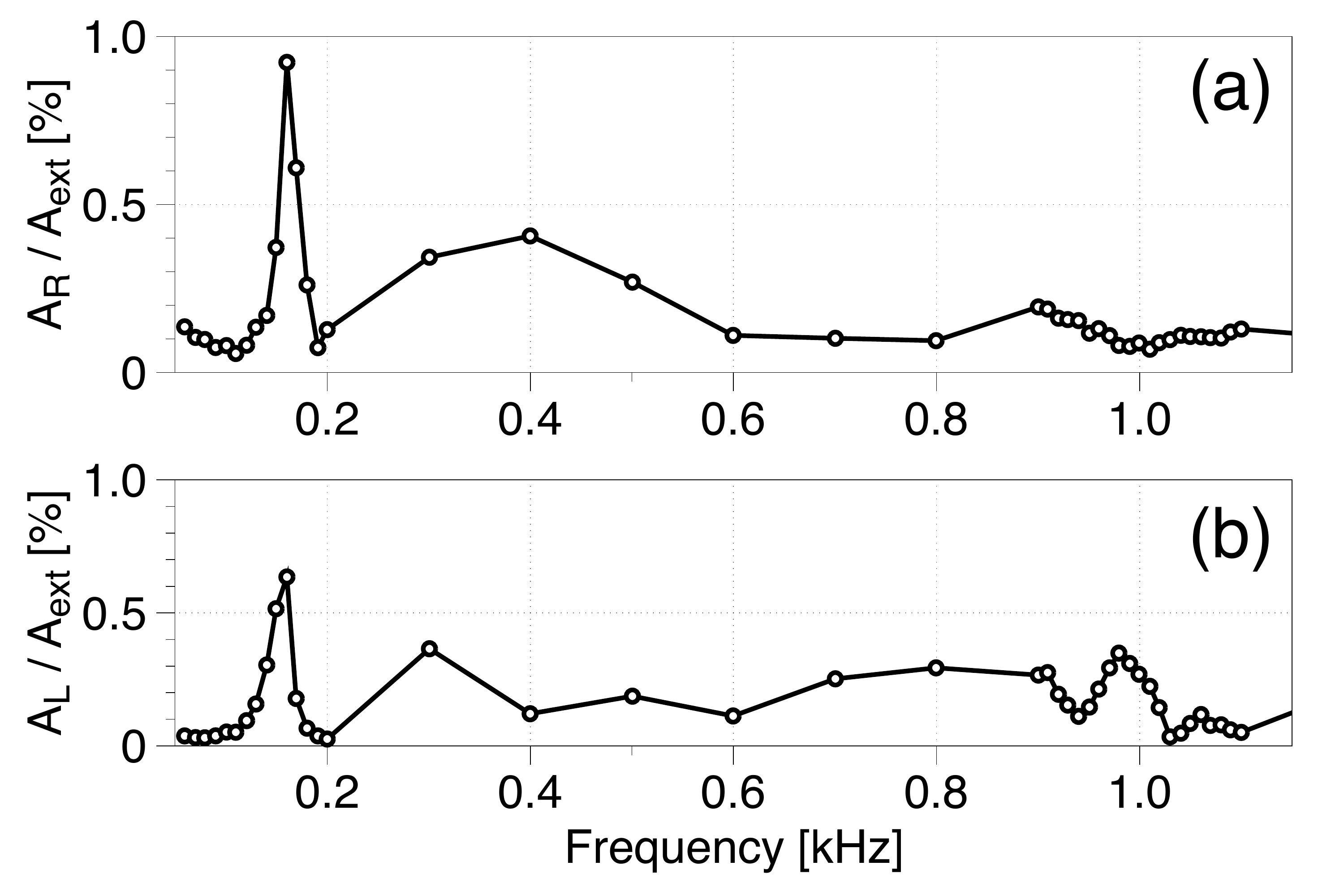}
		\caption{\label{fig:result_resonantmodes} (a) Acceleration amplitude in the radial direction, normalized with respect to the excitation amplitude for different sinusoidal driving frequencies. (b) Normalized longitudinal component as a function of the driving frequency.}
	\end{center}
\end{figure}

\begin{figure}[t]
	\begin{center}
		\includegraphics[scale=0.25]{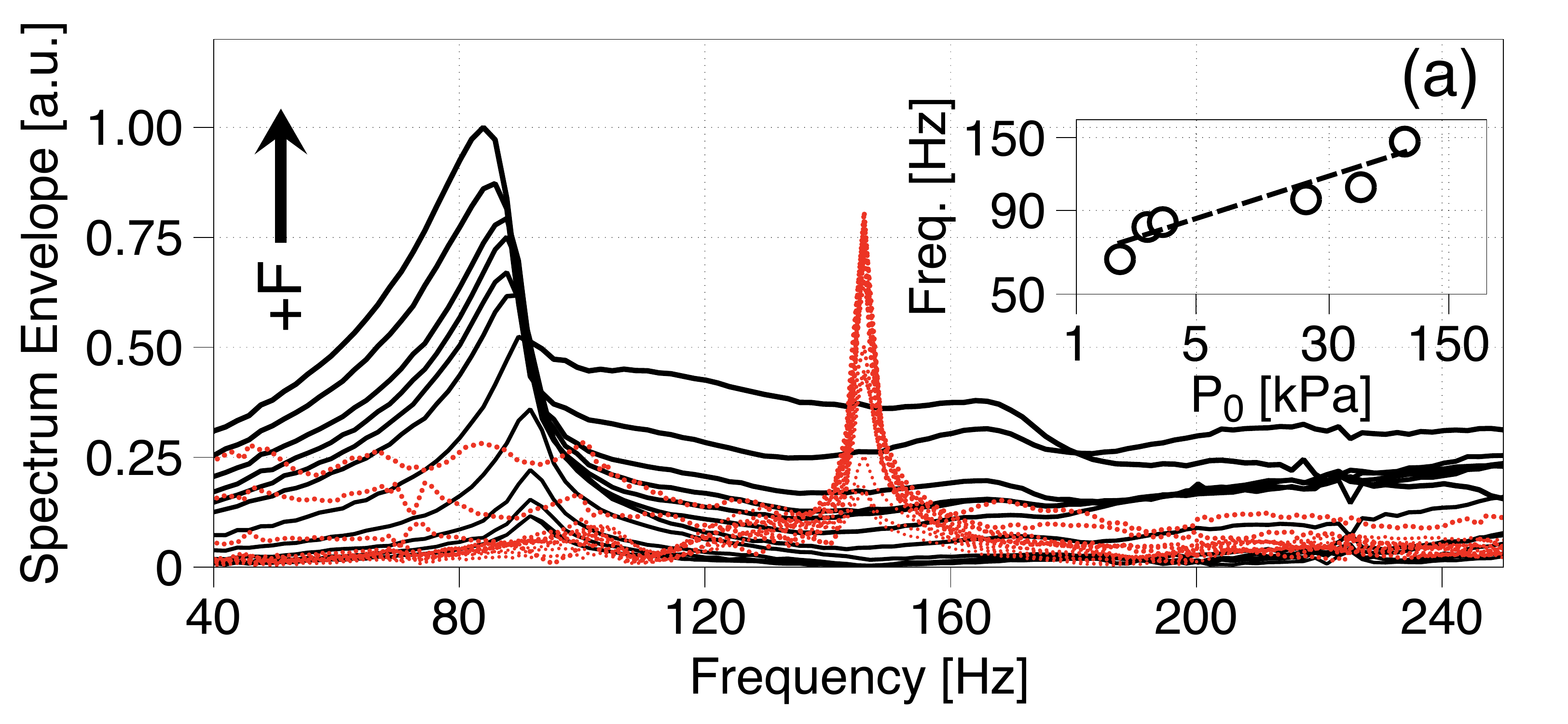}
		\caption{\label{fig:result_envelope}(Color online) (a) Envelopes of the power spectrum of the radial acceleration signal for increasing drive amplitudes, at $P_0=3.2$ kPa (continuous black lines) and $P_0=83$ kPa (dotted red lines). Inset:  Frequency of the radial mode as a function of the static confinement pressure $P_0$; the dashed line corresponds to a $1/6$ slope.}
	\end{center}
\end{figure}

To understand the frequency content of the propagated signals, we investigate the resonant modes of the granular packing by imposing a continuous excitation on the sample.   We couple a vibration exciter (Bruel \& Krajer model 4809 with its corresponding power amplifier) to one end of the cylinder and measure the longitudinal acceleration amplitude ($A_L$) at the opposite end.  The radial component ($A_R$) of the vibration is recorded near the middle of the cylinder mantle by means of an additional accelerometer.  By varying the frequency of excitation, we show that a radial mode appears close to $100$ Hz (Fig.~\ref{fig:result_resonantmodes}). We find no significant longitudinal mode at the amplitude scale of the vibration exciter.
Next, we test the nature of the radial resonance by increasing the amplitude of the sinusoidal driving signal. In Fig.~\ref{fig:result_envelope}, we plot the envelopes of the power spectrum of the radial accelerometer signal for two values of the confining pressure. We observe that the fundamental frequency of the radial mode for low confinement (solid black line) decreases as the wave amplitude increases, resulting in a softer medium at higher amplitudes. However, we do not observe the same behavior for the higher confining pressure (dotted red line), in which case the fundamental frequency does not change with the impulse amplitude.  \\
The diminution of the frequency of a mode with increasing amplitude has been observed in several systems~\cite{prlEspindola,Jia_softening,PAJohnson_nature,PAJohnson_vibration}, and it has been attributed to nonlinear effects arising from the asymmetry of the mechanical response of the system.  At a low confinement pressure and high amplitude excitation, the system is softer in tension and harder in compression.
In addition, we observe that the radial resonant mode increases in frequency with increasing $P_0$. In the insert of Fig.~\ref{fig:result_envelope}, we plot the frequency of this mode for various values of $P_0$, observing that the frequency increase is consistent with a power law of the confinement pressure with an exponent of $1/6$.   This scaling suggests that there is a simple dispersion relation involving the sound speed that is a function of $P_0^{1/6}$ and a geometrical scaling related to the sample diameter; thus, $f_{0} \propto P_0^{1/6}$.\\
Moreover, we observe that this low-frequency peak is wider for a low confinement pressure and narrows as confinement increases, indicating that it is strongly affected by a dissipation that decreases with increasing $P_0$ (see the next section).  This dissipation is expected to be dominated by grain-grain friction at low $P_0$ and low-frequency excitation.  However, as shown in the following section, at the confinements and excitations explored in this work, the viscous effect of the material is the primary mechanism of energy dissipation~\cite{kuwabara_kono}.\\

\subsection{\label{subsec:Attenuation}Attenuation}
\begin{figure}[h!]
	\begin{center}
		\includegraphics[scale=0.25]{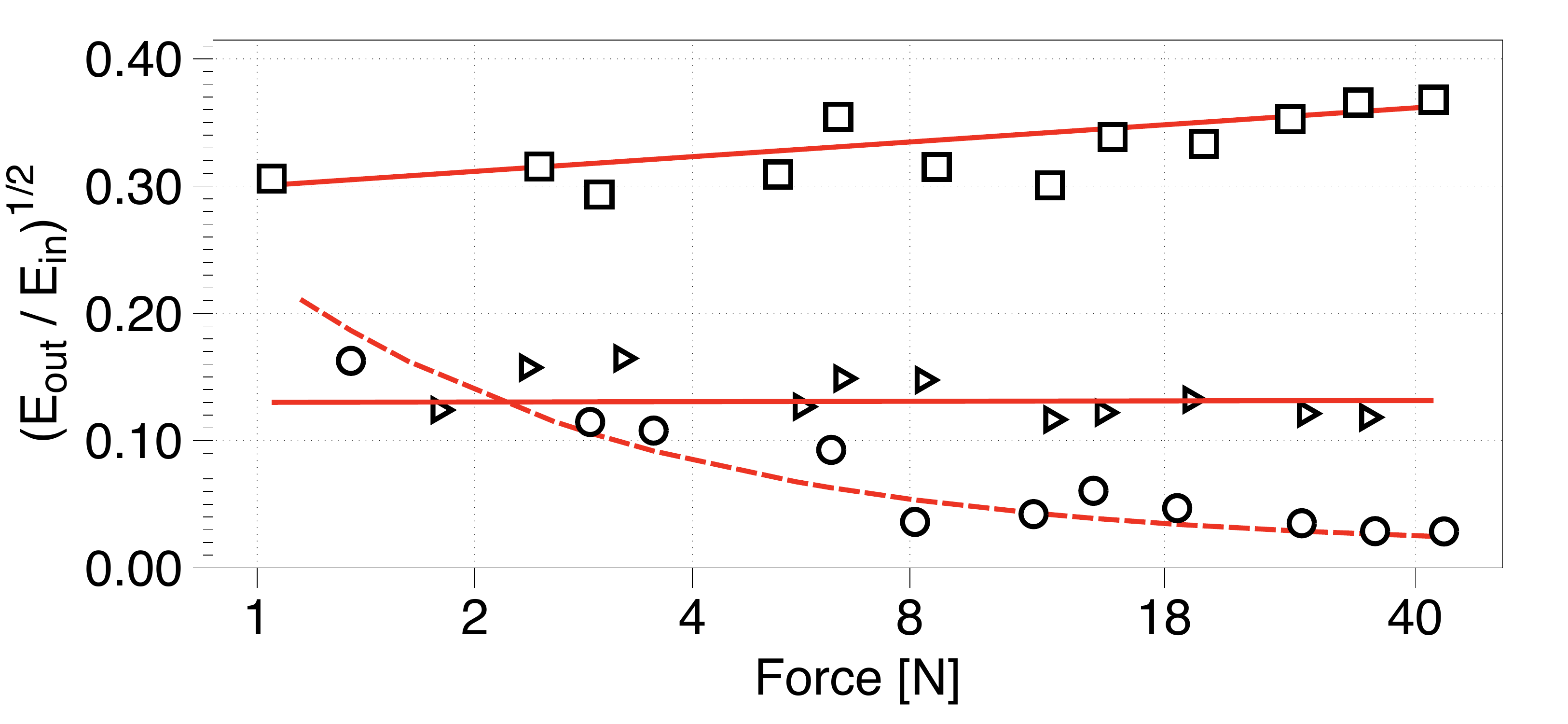}
		\caption{\label{fig:attenuation}(Color online) Square root of the ratio between the outgoing energy $\tilde{E}_{out}$ and the input energy $\tilde{E}_{in}$ as a function of the input force for different values of the confining pressure $P_0$.  The dashed red line corresponds to the LC limit, given in Eq.~(\ref{eq:epsilontotal_LC}), whereas the solid red line corresponds to the HC limit, given in Eq.~(\ref{eq:dissipation}). The symbols are defined as follows: $\circ$: $P_0 =0.5$ kPa, $\triangleright$: $P_0 =23.1$ kPa, $\square$: $P_0=50.0$ kPa.}
	\end{center}
\end{figure}

The geometric features of the boundary conditions in our experiments, i.e., the lack of rigid ends at both cylindrical caps, permit the measurement of the attenuation of the pulse amplitude after propagation through the compact medium.  The ratio of the output energy to the input energy, $\tilde{E}_{out}/\tilde{E}_{in}$, is estimated from the time averages of the force and acceleration at the input and output boundaries, respectively.  To model the energy dissipation, we begin by considering the energy lost between the $n$-th and $n+1$-th layers of the material in the low-confinement (LC) case.   The difference between the maximum potential energies $U$ in these layers is the the work done by the dissipative force at the layer contacts.   In the nonlinear propagation of pulses along linear chains of spheres consisting of a variety of materials, it has been shown that the dissipation is dominated by viscoelasticity~\cite{Job_Melo_Solitary_waves,kuwabara_kono}.  Thus, we write
\begin{equation}
U_{n} - U_{n+1} = \delta \eta \kappa \partial_t \delta^{3/2},
\label{eq:energy_diff}
\end{equation}
where $\eta$ is a material constant. The elastic potential energy is $U_n=(2/5)\kappa \delta^{5/2}$, and the total force is $F=\kappa \delta^{3/2}$, where $\kappa^{-1}=(2\theta)(2/R)^{1/2}$ and $\theta = 3(1-\nu^2)/4Y$ ($\nu$ and $Y$ are the Poisson ratio and Young's modulus, respectively, of the material). Thus, the local restitution coefficient in the LC case, $\epsilon_{LC}$, can be written as follows:
\begin{equation}
1-\epsilon_{LC} \approx \frac{5}{4}\eta \kappa^{2/3} F^{1/6}.
\label{eq:epsilonLC}
\end{equation}
We calculate the force propagated from layer to layer as
\begin{equation}
F_{n+1}=(1-\epsilon_{LC})F_{n}
\label{eq:force_nn1}
\end{equation}
and use the expression obtained in Eq.~(\ref{eq:epsilonLC}) to write the  continuum differential equation: $\frac{dF}{dx}= -C_{\eta}F^{7/6}$, whose solution is written as $F^{1/6}(L)-F^{1/6}(0)=C_{\eta}L/12R$, where $0$ and $L$ indicate the positions of the input and output sensors, respectively.  Then, this result can be combined with Eq.~(\ref{eq:energy_diff}) to obtain the restitution coefficient in the LC limit,
\begin{equation}
\left(\frac{E_{out}}{E_{in}}\right)^{1/2}=\left(\frac{U(L)}{U(0)}\right)^{1/2} = \frac{1}{(1+\frac{C_{\eta}L}{12R}F(0)^{1/6})^{5}},
\label{eq:epsilontotal_LC}
\end{equation}
which leads to $\left(\frac{E_{out}}{E_{in}}\right)^{1/2} \approx CF(0)^{-5/6}$ in the high-amplitude LC limit. The dashed red line in Fig.~\ref{fig:attenuation} shows the fit result, $\left(E_{out}/E_{in}\right)^{1/2} \approx 0.23F(0)^{-5/6}$, from which we extract the value of the characteristic viscoelastic relaxation time for glass, $\eta \approx 0.041 \mu$ s.

Meanwhile, for the high-confinement (HC) limit, we begin by taking the linear limit of the dynamic equation for the $n$-th grain in a 1D chain with viscoelastic coupling~\cite{kuwabara_kono}:
\begin{equation}
m\partial_{tt} u_n = \kappa \delta^{3/2}_{+} - \kappa \delta^{3/2}_{-} + \eta \kappa \partial_t \delta^{3/2}_{+} - \eta \kappa \partial_t \delta^{3/2}_{-}, 
\label{eq:newton_chain}
\end{equation}
where $\partial_{t}$ and $\partial_{tt}$ are the first- and second-order time derivatives, respectively, and $\delta_{+}=\delta_0 + \tilde{\delta}_{+}$ is the total deformation between the $n$-th and $(n+1)$-th grains ($\delta_{-}$ is the corresponding deformation with respect to the $(n-1)$-th grain). We assume that the confinement deformation satisfies $\delta_0 \gg \tilde{\delta}$ and take the Taylor approximation for each term in Eq.~(\ref{eq:newton_chain}).\\
In the continuum limit, we take $u_n=u(x,t)$ and $(\tilde{\delta}_+ - \tilde{\delta}_-)/(4R)^2=\partial_{xx} u(x,t)$ as a second-order spatial derivative. Then, Eq.~(\ref{eq:newton_chain}) becomes a continuous linear wave equation with a dissipative term:
\begin{eqnarray}
\partial_{tt} u(x,t) &=& c_0^2\partial_{xx} u(x,t) + \eta c_0^2 \partial_t \partial_{xx} u(x,t),\\
c_0^2 &=& \frac{9}{2}\frac{\kappa \delta_0^{1/2}}{\pi R \rho_0}.
\label{eq:c02}
\end{eqnarray} 
By assuming a traveling wave solution of the form $u(x,t)=u_0 e^{i(kx-\omega t)}$, we find that
\begin{equation}
k = \frac{\omega}{c_0}(1+\eta^2\omega^2)^{-1/4}e^{i\phi/2}, \\
\label{eq:numero_onda}
\end{equation}
where $\phi=\arctan{(\eta \omega)}$. Given that $\eta\approx10^{-8}$ s and that the range of frequencies experimentally observed in our system is relatively low ($f<10^4$ Hz), $\eta \omega<<1$, such that $\arctan{(\phi)}\approx \eta \omega$.  Then, the real part of $k$, which corresponds to the dispersion relation, yields the sound velocity $c_0= c_{v} P_0^{1/6}$, where we have taken the force at the contact with a bead to be equal to the area of the sphere multiplied by the hydrostatic pressure $P_0$ applied to the packing. In our case, the constant $c_{v}\approx 100$Pa$^{-1/6}$ m/s is a function of the material properties of the beads. 
We calculate the attenuation of the wave as follows:
\begin{equation}
\left(\frac{E_{out}}{E_{in}}\right)^{1/2} = \frac{\partial_t u(x,t)\vline_{x=L}}{\partial_t u(x,t)\vline_{x=0}} \approx e^{-\frac{\eta \omega^2}{2c_0} L} \approx 1 - \frac{\eta \omega^2}{2c_vP_0^{1/6}}L. 
\label{eq:dissipation}
\end{equation} 
As shown in Fig.~\ref{fig:attenuation} for high $P_0$, the restitution coefficient is independent of the impulse amplitude and increases with increasing confinement pressure, whereas it exhibits a scaling of $F^{-5/6}$ for a low confinement pressure. \\

\section{\label{sec:conclusion} Conclusions}
Through experimental evidence, we verified that the propagation of mechanical impulses in granular media strongly depends on both the confinement pressure applied to the packing and the amplitude of the impulse. We tested the dependence of the propagation speed as a function of the static confinement pressure ($P_0$) and the dynamic pulse amplitude ($F$), and we found that the impulse velocity follows a power law proportional to $F^{1/6}$ in the low-confinement case, consistent with the nonlinear propagation described by \cite{dynamics_heterogeneous_nesterenko2001}. Moreover, the impulse velocity becomes independent of the pulse amplitude in the high-confinement case, in which the velocity dependence follows a power law of $P_0^{1/6}$, also consistent with previous experimental evidence. In addition, we studied the resonant modes of the packing by imposing a continuous sinusoidal input driving signal and measuring the radial and longitudinal output accelerations; we found no distinguishable longitudinal mode up to the amplitude limit of the driver, whereas a radial mode was observed that weakened with increasing driving amplitude at low confinement pressures and became independent of the excitation at high confinement. Consistently, the frequency of the radial mode increased with increasing $P_0$, with the same dependence as that of the sound speed. Finally, we tested the nature of the dissipation mechanism using a viscoelastic model \citep{kuwabara_kono} in both the high- and low-confinement conditions. We first used a ballistic approximation to calculate the restitution coefficient for a low confinement pressure and obtained a power-law dependence of the coefficient, in agreement with our experimental evidence. Finally, in the high-confinement case, we linearized the dynamic 1D equation for a chain of grains to obtain the propagation velocity and the dissipation of the propagating wave, thereby showing that the dissipation is independent of the wave amplitude and that it increases rapidly with $\omega^2$ and decreases slowly with $P_0^{-1/6}$.

{\bf Acknowledgments:}\\
F. Santibanez acknowledges the financial support from FONDECYT Project No.11140556 and the PUCV-VRIEA grant DI-Iniciaci\'on  No. 37.379/2014.   F. Melo is extremely grateful to the NICOP Program of the US Navy for additional support.

\bibliographystyle{unsrt}
\bibliography{biblio.bib}

\end{document}